# BOUNDS ON THE SCALE OF NONCOMMUTATIVITY FROM MONO PHOTON PRODUCTION IN ATLAS RUNS -1 AND -2 EXPERIMENTS AT LHC ENERGIES

MOHAMED REDA BEKLI

*Laboratoire de Physique Théorique, Faculté des Sciences de la Nature et de la Vie, Université de Bejaia, 06000 Bejaia, Algeria*
*astronomes@gmail.com*

ILHEM CHADOU

*Département de Technologie, Faculté de Technologie, Université de Bejaia, 06000 Bejaia, Algeria*
*ilhem_work@yahoo.fr*

NOUREDDINE MEBARKI

*Laboratoire de Physique Mathématique et Subatomique, Mentouri University, Constantine, Algeria*
*nnmebarki@yahoo.fr*

Leading order study of direct photon production from proton-proton collisions, in the framework of Minimal (Seiberg-Witten) Non-Commutative Standard Model (NCSM), taking into account the Earth-rotation effects.

We found that relative non-commutative contributions increase significantly at very high photon transverse momentum. Therefore, using Run-1 ($\sqrt{s}$= 8 TeV) and Run-2 ($\sqrt{s}$= 13 TeV) ATLAS experimental data of inclusive isolated prompt photon cross-section, TeV-Scale bounds of the non-commutativity (NC) parameter are obtained.

For space-space non-commutativity, we obtain: $\Lambda_b$=1.145±0.015 TeV , and for space-time non-commutativity, we obtain : $\Lambda_e$=1.125±0.035 TeV.

## 1. Introduction

The main objective of modern physics is to describe in a coherent and unified way the behavior of the matter at the microscopic level and try to understand how these particles interact to form the ordinary macroscopic matter and explain all observable phenomena. Not being completely satisfied by the current standard model of particle physics (SM), many other extensions beyond SM have been developed so far. Indeed, despite its success, the SM has some deficiencies and many issues remain unexplained leading to the developments of more satisfactory alternative models. Moreover, the LHC experiments performed in CERN at high energies and luminosities gives us the opportunity to test some of them. In this paper, we explore one of those theories the so-called minimal non-commutative standard model (Minimal-NCSM), an alternative theory where its phenomenological implications and confrontations with the LHC experimental results are under investigation by many physicists during the last few years. It is worth to mention that the NCSM is based essentially on the space-time non-commutativity and is constructed using the Moyal-Weyl product and Seiberg-Witten (SW) Maps [1].

Lower-bounds on the NC parameter within the NCSM has been the subject of several studies. In fact, in the context of high energies particle physics, several constraints were obtained by many authors using the various collider experimental data such as LEP [2], LHC [3] and Tevatron [4]. Other experimental bounds and projected sensitivities for the future colliders are discussed in [5]. It should be noted that a very high bounds are obtained for several constructions of the non-commutative theories in ultra-precision experiments at low energies and astrophysical systems.

In the present paper, we investigate (in the context of the minimal NCSM constructed using Moyal-Weyl star product and SW Maps) the most clean process to test perturbative QCD predictions in hadronic collisions which is the prompt photon production in proton-proton collisions at the LHC energies (Run-1 and Run-2). This process involves novel tree level contributions arising from the Minimal-NCSM, not considered before. The calculations of the corresponding inclusive cross section are performed by taking into account the Earth-rotation effects. Our results confronted with the ATLAS experimental data of the inclusive isolated prompt photon cross section, recorded at a center-of-mass energy of √s=8 TeV and √s=13 TeV, allowing us to deduce a new TeV scale lower bounds on the NC type space-space and space-time parameter ($\Lambda^{bound}$).

The paper is organized as follows: in §2 we described the theoretical model (Minimal-NCSM) that has been the subject of numerous publications, in §3 and §4 we discuss the direct photon production within the NCSM taking into account the Earth-rotation effects, in §5 we present and discuss the numerical results, in §6 we draw our conclusions. Finally, the cross section calculations are described in appendix A.

## 2. Theoretical framework

The non-commutative space and time can be realized in terms of coordinate operators $\hat{x}^\mu$ satisfying the following relation:

$$[\hat{x}^\mu, \hat{x}^\nu] = i\Theta^{\mu\nu} \tag{1}$$

In what follows, we set $\Theta$ as a real antisymmetric and constant matrix.

The action for field theories is obtained using SW Maps [1] and after replacing all usual products by the Moyal-Weyl product (or star product) defined by the following power series expansion:

$$(f * g)(x) = exp\left(\frac{i}{2}\Theta^{\mu\nu}\frac{\partial}{\partial x^\mu}\frac{\partial}{\partial y^\nu}\right)f(x)g(y)\bigg|_{y\to x} \tag{2}$$

$f(x)$ and $g(y)$ denote two any functions on $\mathbb{R}$, an auxiliary space used to define Moyal-Weyl product.

This method makes it possible to construct an equivalent model of quantum field theory that is function of commutative space and time coordinates. A first version of the non-Abelian Yang–Mills theory on non-commutative space and time is presented in [6, 7, 8].

The different choices for representations of the gauge group yield two versions of the Non-Commutative Standard Model: Minimal and non-Minimal NCSM. For a detailed description of this model, constructed on $SU(3)_C \otimes SU(2)_L \otimes U(1)_Y$ group in first order of $\Theta$, see [9, 10, 11]. Initially, the extension of the NCSM in second order of the non-commutative parameter shows many ambiguities [12], and make the calculations indefinite. However, over the last 2 decades θ-exact SW map was computed [13, 14], and θ-exact NC-QED model constructed. Analyzing 1-loop photon 2-point functions the SW map freedom parameters of the model were all uniquely fixed [15, 16]. In that sense minimal θ-exact SW map NCSM is possible and it was constructed in [17].

An important effect emerges from non-commutative gauge theories is the so-called ultraviolet/infrared (UV/IR) mixing that is absent in ordinary QFT [18, 19]. This property, also present on a κ-deformed Euclidean space for a real scalar $\phi^4$ theory [20] and in Nonassociative Snyder $\phi^4$ theory [21], has been the subject of many studies in the framework of the NCQFT with SW Maps. Thus, in [13] the fermion contribution of the one loop correction to the photon propagator is computed and it is found that it gives the same UV/IR mixing term as in NC-QED without SW map. In [16, 22, 23], it was found that NCQFT based on θ-exact formulation improve UV/IR behavior in its supersymmetric version.

It is important to note that the unitarity and causality of the NC-QFT (non-commutative quantum field theory) are safe for space-like non-commutativity, while for the time-like non-commutativity they are broken [24, 25, 26]. However, the time-like NC could be replaced with the light-like (where $\theta^{0i} = -\theta^{1i}, \forall\, i = 1,2,3$ [25]) unitary safe NC-QFT (also called perturbative unitarity [26]), belonging to the class of Very Special Theory of Relativity [27]. See for example the analysis of the Z boson decays ($Z \to \gamma\gamma, \nu\bar{\nu}$), for the two unitarity cases [28]: light-like and space-like non-commutativity.

The LHC experiments gives us the opportunity to test different theories beyond standard model. Recently, the ATLAS measurements of light-by-light scattering ($\gamma\gamma \to \gamma\gamma$) in Pb-Pb collisions [29, 30], which is the first direct evidence for this process, imposed a strong constraint on Born-Infeld Theory [31] and on nonlinear Lorentz-violating effects in electrodynamics [32]. These same scattering cross section measurements are used to obtain the most precise constraints for nonlinear corrections to Maxwell electrodynamics [33]. NCQED cross section calculations show that the current ATLAS experiment at a center-of-mass energy of $\sqrt{s_{NN}} = 5.02\, TeV$ can only probe $\Lambda < 100\, GeV$ region, however the next generation hadron colliders (such as the SppC) could have the potential to probe the non-commutative scale up to $\sim 300$ GeV [34].

In the present study, we explore the phenomenological consequences of Minimal-NCSM in PP collisions at LHC energies. Theoretically, additional interaction terms that does not exist in the ordinary standard model must be taken into account. Technically, to calculate the transition amplitude we use the Feynman rules listed in [10, 11]. For direct photon process analyzed here, novel tree level contributions are taken into account.

## 3. Direct photon production

Direct photon production is one of the most effective processes used to probe the structure of hadrons. Indeed, these photons emerge from the hard process and provide information of the parton dynamic inside the hadrons. The kinematics and theoretical framework of this process is presented in [35].

In what follows, we focus on the production of direct photon from proton-proton collision (pp → γ + X) in the framework of Minimal-NCSM at center-of-mass energies $\sqrt{s}$ = 8 TeV and $\sqrt{s}$ = 13 TeV. At the partonic level, only two subprocesses contribute at lowest order: quark-antiquark annihilation ($q\bar{q} \to g\gamma$, see Fig. 1) and quark-gluon Compton scattering ($qg \to q\gamma$, see Fig. 2). It should be noted that the Feynman diagrams of Fig. 1 (c) and Fig. 2 (c) are purely non-commutative.

The differential cross sections are calculated for each of the two subprocesses. Spin and polarization sums (trace calculations) are performed using computer code with FeynCalc [36, 37]. After that, it is necessary to applicate a series of rotations to express ϴ in the Laboratory Frame. Additional detailed calculations and the complete expressions of differential cross sections are given in §7 (Appendix A. cross section calculations).

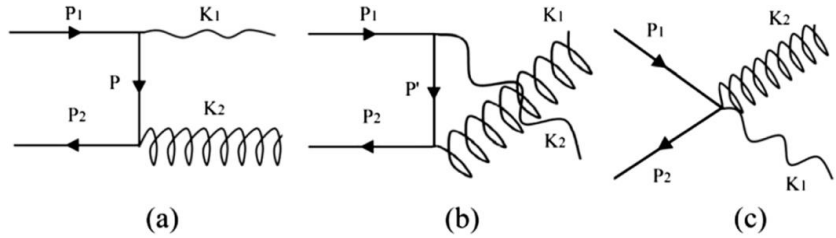

Fig. 1. Feynman diagrams of quark-antiquark annihilation subprocess : $q\bar{q} \to g\gamma$.

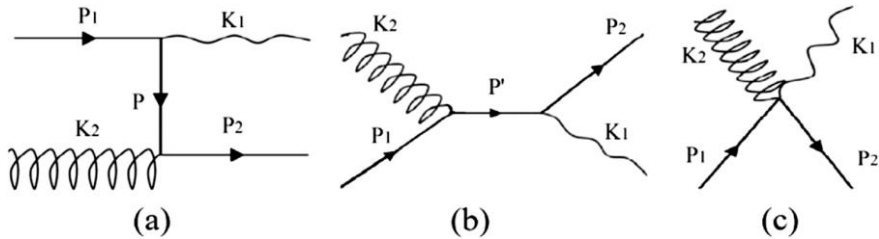

Fig. 2. Feynman diagrams of quark-gluon Compton scattering subprocess : $qg \to q\gamma$.

The inclusive cross section is obtained by an incoherent summation over all the possible sub-processes (see Appendix A.). The calculation is done separately in the two different cases of non-commutativity, space-space and space-time. For Numerical evaluation of the multiple integral, we use Adaptive Monte Carlo Method.

It is important to note that in collider experiments, the other process that contributes to the prompt-photon production is the fragmentation process. However, the application of the isolation criteria in the measurements reduces the parton-to-photon fragmentation contribution to the total cross section. Indeed, the most recent measurements of ATLAS experiment [38, 39], performed with the isolation requirement $P_T^{iso} < 4.8\,\text{GeV} + 4.2\ 10^{-3} P_T^\gamma$ calculated within a cone of size $\Delta R = 0.4$, reduce significantly the fragmentation components, as indicated in [40]. Furthermore, the contributions from the fragmentation decrease considerably when $|\cos\theta^*| \to 0$ [41, 42]. Even without imposing isolation cuts, the Compton process dominates for $p_T^\gamma$ above 45 GeV at midrapidity region ($y = 0$) [43]. Accordingly, it is not necessary to include the fragmentation contribution in our calculations, otherwise the computations become extremely tedious and time consuming; this is the major advantage of the direct photon process.

## 4. Earth-rotation contribution

By convention, we consider that initial particle moves along the $x_3$ axis of the Laboratory Coordinate System (collision beam axis), with θ the scattering angle and φ the transverse angle. The second axis $x_2$ is taken perpendicular to the Earth surface.

On other hand, we chose as initial coordinate system the Celestial Equatorial Coordinate System, totally independent of the Earth's rotation, as in [2]. The third axis $X_3$ is oriented along the Earth rotation axis, and the other two axes following a perpendicular direction, needless to say in our case.

We decompose $\Theta^{\mu\nu}$ matrix into electric-like components $\vec{E} = (\Theta^{01}, \Theta^{02}, \Theta^{03})$ and magnetic-like components $\vec{B} = (\Theta^{23}, \Theta^{31}, \Theta^{12})$ according the two types of non-commutativity, space-time and space-space, respectively. The orientation of these two vectors in the celestial coordinate system is given by the following parameterization:

$$\vec{B} = \frac{1}{\Lambda_b^2}\begin{pmatrix} \cos\beta_b \sin\gamma_b \\ \sin\beta_b \sin\gamma_b \\ \cos\gamma_b \end{pmatrix}, \vec{E} = \frac{1}{\Lambda_e^2}\begin{pmatrix} \cos\beta_e \sin\gamma_e \\ \sin\beta_e \sin\gamma_e \\ \cos\gamma_e \end{pmatrix} \qquad (3)$$

We indicate by β and γ the spherical coordinates of $\vec{E}$ and $\vec{B}$, the azimuthal angle and colatitude, respectively, as illustrated in Fig. 3 (a). $\Lambda_b$ and $\Lambda_e$ are the two independents NC parameter, considered on the present theory as fundamental constants in nature.

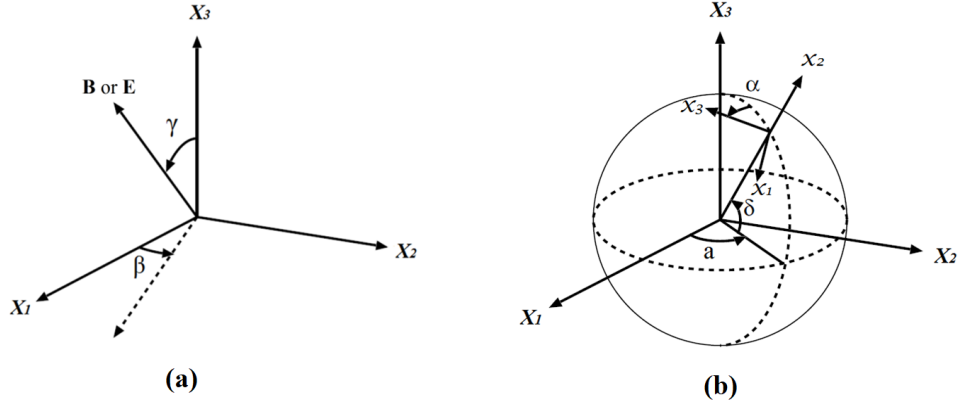

Fig. 3. (a) $\vec{E}$ and $\vec{B}$ in celestial equatorial coordinate system (b) Laboratory coordinate system at latitude= δ and inclination= α

To express $\vec{E}$ and $\vec{B}$ in the laboratory frame of reference located at δ latitude, where the collision beam axis $x_3$ is inclined at an angle α to the North direction (see Fig. 3 (b)), we should execute a series of rotations Eqs. (4) and (5), as in [44, 45].

$$\{\vec{B'}, \vec{E'}\} = R\{\vec{B}, \vec{E}\} \quad (4)$$

Where,

$$\begin{cases} R = R_2(\alpha) R_3\left(-\dfrac{\pi}{2}\right) R_2(-\delta) R_3(a) \\ a = \omega\, t + a_0 \end{cases} \quad (5)$$

The parameter "a" represents the right ascension of the laboratory site, where collisions take place, and ω the angular velocity of the Earth. Indeed, since the collisions occur for several months, we average over β (previously defined), φ (the azimuthal angle of final particles) and the time t (therefore on a). We keep only the colatitudes γ of $\vec{E}$ and $\vec{B}$, which remains fixed during this period.

## 5. Numerical results

The variation of the inclusive cross sections as a function of the transverse momentum are showed in Fig. 4 (a-f) show. The computations are performed for two center-of-mass energies reached at the LHC $\sqrt{s} = 8$ and 13 TeV, and for different values of the NC parameter, using the recent CT14 PDF [46]. As a comparison, we show the ATLAS measurement results of the inclusive isolated prompt photon cross section obtained at a center-of-mass energies of $\sqrt{s} = 8$ TeV based on an integrated luminosity of $L_{int} = 20.2$ fb$^{-1}$ [38] and $\sqrt{s} = 13$ TeV based on $L_{int} = 3.2$ fb$^{-1}$ [39].

We see clearly that for $\Lambda > 1\, TeV$, the non-commutative contribution becomes perceptible at very large $P_T^{\gamma}$. Thus, the lower bounds on the NC parameter can be obtained at high $P_T^\gamma$ for different values of colatitude γ by taking into account the uncertainties (systematic and statistical uncertainties) δσ, using Eq. (6).

$$\left(\frac{d\sigma}{dp_T^\gamma}\right)_{NC}(\Lambda^{bound}) = \left(\frac{d\sigma}{dp_T^\gamma}\right)_{exp} + \Delta\left(\frac{d\sigma}{dp_T^\gamma}\right) \qquad (6)$$

The bounds are computed separately for each cases of non-commutativity: space-space ($\Lambda_b^{bound}$) and space-time ($\Lambda_e^{bound}$). The greatest values of $\Lambda^{bound}$ are obtained at $P_T^\gamma \in [1100,1500]\ GeV$ where the error bar is the most restrictive. To avoid any overestimation, we take $P_T^\gamma = 1500\ GeV$ to obtain an absolute lower bound.

A careful analysis of the analytic expression of the two partonic cross sections (Compton and Annihilation) shows that all terms that depend on γ vary as a function of $\cos 2\gamma$. This is attributable to the fact we have averaged over β, the azimuthal angle that appears in Eq. (3). thus, it is possible to restrict the calculations for γ varying from 0° to 90°.

The bounds listed in Table 1 are calculated using the results of the Atlas collaboration obtained at a center-of-mass energy of $\sqrt{s} = 8$ TeV for the photon pseudorapidity region [38]: $|\eta| < 0.6$. Thus, for space-space non-commutativity, we obtain $\Lambda_b^{bound} \in [0.96,0.97]$ TeV, and for space-time non-commutativity, we obtain $\Lambda_e^{bound} \in [0.93,0.99]$ TeV, for $\gamma_{b,e} \in [0°, 90°]$.

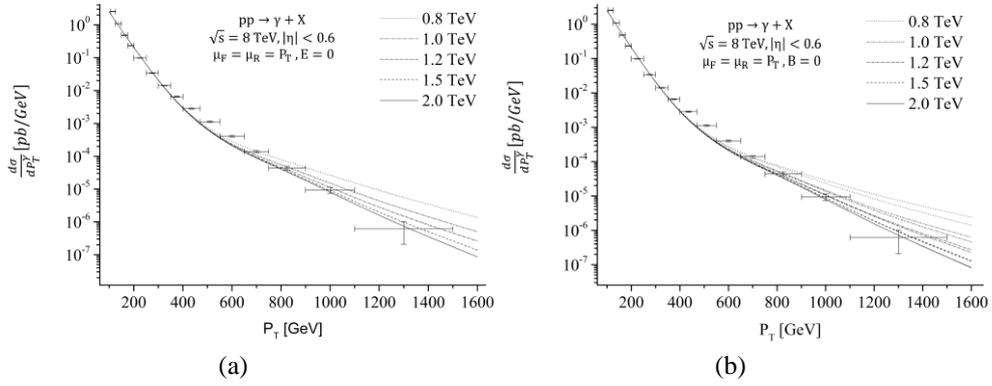

(a)     (b)

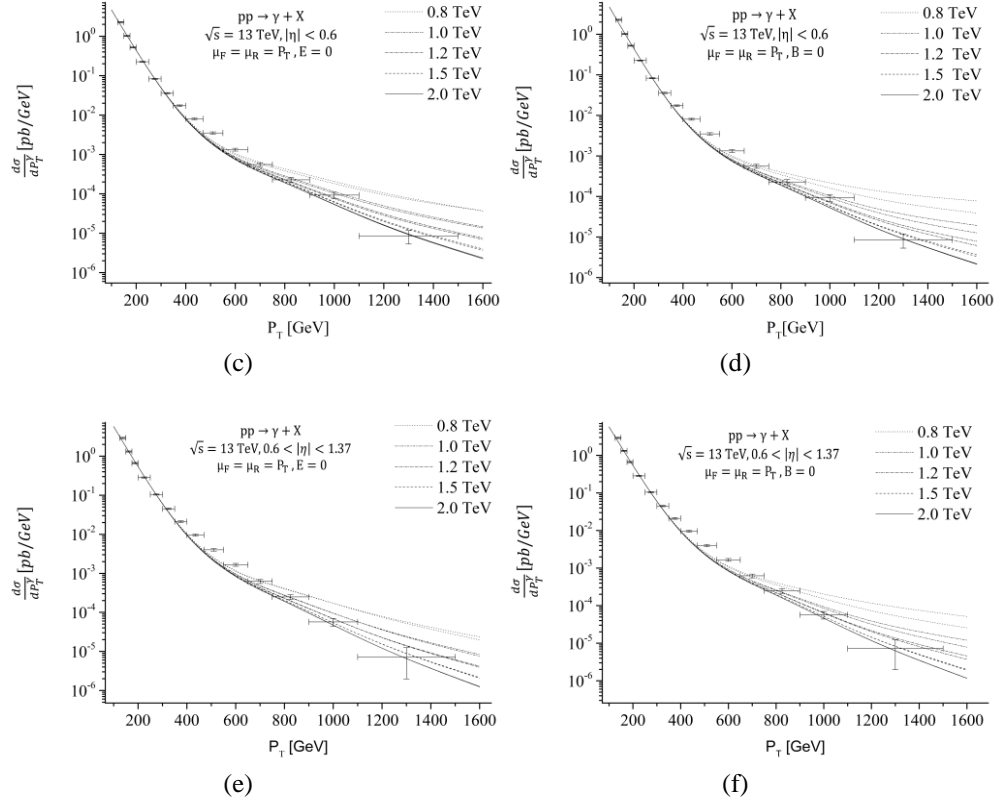

**Fig. 4.** Non-commutative inclusive cross section as a function of the photon transverse momentum, for non-commutativity parameter varying between 08 to 2.0 TeV and a colatitude γ = 0° and 90∘. The ATLAS measurements obtained at a center-of-mass energy of $\sqrt{s}$ = 8 TeV [38] and $\sqrt{s}$ = 13 TeV [39] are also represented. On the whole, the cross-section for $\gamma = 90°$ is greater than the cross-section for $\gamma = 0°$, except in (a) and (e).

Table 1 Lower bounds on the non-commutativity parameter obtained at $\sqrt{s} = 8\ TeV$.

| γ | Space-Space : $\Lambda_b^{bound}$ [TeV] | Space-Time : $\Lambda_e^{bound}$ [TeV] |
|---|---|---|
| 0° | 0.96 | 0.93 |
| 30° | 0.96 | 0.96 |
| 60° | 0.96 | 0.99 |
| 90° | 0.97 | 0.99 |

The bounds listed in Table 2 are calculated using the results of the Atlas collaboration obtained at a center-of-mass energy of $\sqrt{s} = 13\,TeV$ for the two regions of photon pseudorapidity [39] : $|\eta| < 0.6$ and $0.6 < |\eta| < 1.37$.

In the case of space-space non-commutativity, we obtain: $\Lambda_b \in [1.13, 1.16]$ TeV for $|\eta| < 0.6$ and $\Lambda_b \in [0.99, 0.97]$ TeV for $0.6 < |\eta| < 1.37$. In the case of space-time non-commutativity, we obtain: $\Lambda_e \in [1.09, 1.16]$ TeV for $|\eta| < 0.6$ and $\Lambda_e \in [0.97, 1.04]$ TeV for $0.6 < |\eta| < 1.37$.

Table 2 Lower bounds on the non-commutativity parameter obtained at $\sqrt{s} = 13\,TeV$.

| $\gamma$ | Space-Space : $\Lambda_b^{bound}$ [TeV] | | Space-Time : $\Lambda_e^{bound}$ [TeV] | |
|---|---|---|---|---|
| | $|\eta| < 0.6$ | $0.6 < |\eta| < 1.37$ | $|\eta| < 0.6$ | $0.6 < |\eta| < 1.37$ |
| 0° | 1.13 | 0.99 | 1.09 | 0.97 |
| 30° | 1.14 | 0.98 | 1.11 | 1.00 |
| 60° | 1.15 | 0.97 | 1.15 | 1.03 |
| 90° | 1.16 | 0.97 | 1.16 | 1.04 |

In most cases, the largest values of $\Lambda^{bound}$ are obtained for $\gamma = 90°$, where non-commutative effects are greater. It should also be noted that the bounds obtained at $\sqrt{s} = 8\,TeV$ are slightly lower than the bounds obtained at $\sqrt{s} = 13\,TeV$ for $|\eta| < 0.6$, even with the same transverse momentum scale $1100 \leq P_T \leq 1500$ GeV. This is not attributable to the difference of center-of-mass energies, but to the fact that the statistical uncertainty in [38] of 62.1% is much higher than the statistical uncertainty of [39], whose value is 33.2%.

Further calculations, not reported here, show that the $\Lambda^{bound}$ obtained for $\sqrt{s} = 7\,TeV$ are much lower than 1 TeV. This is due to the relatively low value of the transverse momentum, which does not exceed 1000 GeV for ATLAS [47, 48, 49], and 400 GeV for CMS [50, 51]. It is therefore not necessary to include them in the present paper. Furthermore, we confronted our leading-order model with recent ATLAS measurements of isolated photon productions in association with a b-jet and a c-jet ($\gamma + b$ and $\gamma + c$) at $\sqrt{s} = 8\,TeV$ [52], yet the results have not come up to our expectations.

## 6. Conclusion

In this paper, we have calculated the inclusive cross-section of direct photon production from proton-proton collisions at Run-1 and Run-2 LHC energies ($\sqrt{s} = 8$ and $13\,TeV$), in the framework of the Minimal Non-Commutative Standard Model, constructed with Moyal-Weyl product and Seiberg-Witten Map. To take partially into account the Earth-rotation effect, we have averaged over all the angles not accessible experimentally, with the exception of the diffusion angle θ and the colatitude γ of $\vec{E}$ and $\vec{B}$.

We found that the non-commutative effects increase considerably with photon transverse energy, which allowed us to deduce TeV-scale bounds on the NC parameter ($\Lambda^{bound}$) using the Data of ATLAS experiment of the LHC. These limits are calculated for colatitude γ varying from 0 ° to 90°.

For a center-of-mass energy of $\sqrt{s} = 8$ TeV and photon pseudorapidity region $|\eta| < 0.6$, we obtain: In the case of a space-space non-commutativity : $\Lambda_b \in [0.96, 0.97]$ TeV, and in the case of a space-time commutativity : $\Lambda_e \in [0.93, 0.99]$ TeV.

For a center-of-mass energy of $\sqrt{s} = 13\ TeV$, we obtain: In the case of a space-space non-commutativity : $\Lambda_b \in [1.13, 1.16]$ TeV for $|\eta| < 0.6$ and $\Lambda_b \in [0.99, 0.97]$TeV for $0.6 < |\eta| < 1.37$. In the case of a space-time non-commutativity: $\Lambda_e \in [1.09, 1.16]$TeV for $|\eta| < 0.6$ and $\Lambda_e \in [0.97, 1.04]$TeV for $0.6 < |\eta| < 1.37$.

In summary, the most important limits obtained in this paper are: $\Lambda_b^{bound} = 1.145 \pm 0.015$ TeV for a space-space non-commutativity, and $\Lambda_e^{bound} = 1.125 \pm 0.035$ TeV for a space-time non-commutativity. Here, the uncertainty is defined as the biggest deviation of $\Lambda^{bound}(\gamma)$ from the mean value.

The main challenge is to extend the calculations at the next-to-leading-order, but it is not easy at all because of the large number of Feynman diagrams needed with a very complex interactions.

**Appendix A. cross section calculations**

In what follows, we decompose $\Theta^{\mu\nu}$ matrix into electric-like components $\vec{E} = (\Theta^{01}, \Theta^{02}, \Theta^{03}) = (E_1, E_2, E_3)$ and magnetic-like components $\vec{B} = (\Theta^{23}, \Theta^{31}, \Theta^{12}) = (B_1, B_2, B_3)$ according to the two types of non-commutativity, space-time and space-space, respectively. We put $E_{ij}$ and $B_{ij}$, as $E_{ij} = \sqrt{E_i^2 + E_j^2}$ and $B_{ij} = \sqrt{B_i^2 + B_j^2}$. The orientation of $\vec{E}$ and $\vec{B}$ in the celestial coordinate system are represented in Fig. 3 (a), with parameterization given by Eq. (3).

By convention, we consider $x_3$ axis of the Laboratory Coordinate System as the beam axis, inclined at an angle α to the North direction (see Fig. 3 (b)), with θ the scattering angle and φ the transverse angle defined with respect to the $x_1$ axis.

For quark-antiquark annihilation subprocess (q$\bar{q}$ → gγ), represented by the diagrams of Fig. 1, the transition amplitudes are given by the following expressions obtained using the Feynman rules listed in [10, 11] :

$$M_A^a = -ieqg_s \bar{v}^{(s)}(2)(c_2^+)_i^m \left[\gamma_\mu - \frac{i}{2}(-K_2)^\nu\left(\Theta_{\mu\nu\rho}P^\rho - \Theta_{\mu\nu}M_q\right)\right]\left(\frac{\lambda^a}{2}\right)_{ij}$$
$$\frac{\not{P} + M_q}{P^2 - M_q^2}\left[\gamma_\alpha - \frac{i}{2}(-K_1)^\beta\left(\Theta_{\alpha\beta\lambda}P_1^\lambda - \Theta_{\alpha\beta}M_q\right)\right]u^{(s')}(1)(c_1^+)_j^n\left(\varepsilon_{2\mu}^* a_2^{al}\right)\varepsilon_{1\alpha}^*$$
(A.1)

$$M_A^b = -ieqg_s \bar{v}^{(s)}(2)(c_2^+)_i^m \left[\gamma_\mu - \frac{i}{2}(-K_1)^\nu\left(\Theta_{\mu\nu\rho}P'^\rho - \Theta_{\mu\nu}M_q\right)\right]\frac{\not{P}' + M_q}{P'^2 - M_q^2}$$
$$\left[\gamma_\alpha - \frac{i}{2}(-K_2)^\beta\left(\Theta_{\alpha\beta\lambda}P_1^\lambda - \Theta_{\alpha\beta}M_q\right)\right]\left(\frac{\lambda^a}{2}\right)_{ij}u^{(s')}(1)(c_1^+)_j^n\left(\varepsilon_{1\mu}^* a_2^{al}\right)\varepsilon_{2\alpha}^*$$
(A.2)

$$M_A^c = -\frac{1}{2} eqg_s \bar{v}^{(s)}(2)(c_2^+)_i^m \Theta_{\mu\nu\rho} \left(\frac{\lambda^a}{2}\right)_{ij} \left[-(K_1 - K_2)^\rho\right] u^{(s')}(1)(c_1^+)_j^n (\varepsilon_{2\nu}^* a_2^{al}) \varepsilon_{1\mu}^* \qquad (A.3)$$

For the second sub-process, quark-gluon Compton scattering (qg → qγ) represented by the diagrams of Fig. 2, we can write:

$$M_C^a = -ieqg_s \bar{u}^{(s)}(2)(c_2^+)_i^m \left[\gamma_\mu - \frac{i}{2} K_2^{\,\nu} \left(\Theta_{\mu\nu\rho} P^\rho - \Theta_{\mu\nu} M_q\right)\right] \left(\frac{\lambda^a}{2}\right)_{ij}$$

$$\frac{\slashed{P} + M_q}{P^2 - M_q^2} \left[\gamma_\alpha - \frac{i}{2}(-K_1)^\beta \left(\Theta_{\alpha\beta\lambda} P_1^\lambda - \Theta_{\alpha\beta} M_q\right)\right] u^{(s')}(1)(c_1^+)_j^n (\varepsilon_{2\mu} a_2^{al}) \varepsilon_{1\alpha}^* \qquad (A.4)$$

$$M_C^b = -ieqg_s \bar{u}^{(s)}(2)(c_2^+)_i^m \left[\gamma_\mu - \frac{i}{2}(-K_1)^\nu \left(\Theta_{\mu\nu\rho} P'^\rho - \Theta_{\mu\nu} M_q\right)\right] \frac{\slashed{P}' + M_q}{P'^2 - M_q^2}$$

$$\left[\gamma_\alpha - \frac{i}{2} K_2^{\,\beta} \left(\Theta_{\alpha\beta\lambda} P_1^\lambda - \Theta_{\alpha\beta} M_q\right)\right] \left(\frac{\lambda^a}{2}\right)_{ij} u^{(s')}(1)(c_1^+)_j^n (\varepsilon_{2\alpha} a_2^{al}) \varepsilon_{1\mu}^* \qquad (A.5)$$

$$M_C^c = -\frac{1}{2} eqg_s \bar{u}^{(s)}(2)(c_2^+)_i^m \Theta_{\alpha\mu\rho} \left(\frac{\lambda^a}{2}\right)_{ij} \left[-(K_1 - K_2)^\rho\right] u^{(s')}(1)(c_1^+)_j^n (\varepsilon_{2\mu} a_2^{al}) \varepsilon_{1\alpha}^* \qquad (A.6)$$

The vectors "c" of 3 components and "a" of 8 components, are used to describe the quarks and gluons color states, as defined in [53]. The indices A and C, respectively, denote the Annihilation and Compton processes, and "λ" represent the eight Gell-Mann matrices. $\Theta_{\mu\nu\rho}$ is defined as $\Theta_{\mu\nu\rho} = \Theta_{\mu\nu}\gamma_\rho + \Theta_{\nu\rho}\gamma_\mu + \Theta_{\rho\mu}\gamma_\nu$.

Neglecting the mass term, the incoming and outgoing 4-momenta are set as:

$$P_{in} = \frac{\sqrt{\hat{s}}}{2}(1,0,0,\pm 1) \qquad (A.7)$$

$$P_{out} = \left(\frac{\sqrt{\hat{s}}}{2}, \pm \vec{p}_{out}\right) \qquad (A.8)$$

Where,

$$\vec{p}_{out} = \frac{\sqrt{\hat{s}}}{2}(\sin\theta\cos\varphi, \sin\theta\sin\varphi, \cos\theta) \qquad (A.9)$$

φ is the azimuthal angle of final particles.

The differential cross section $d\hat{\sigma}/d\hat{t}$ is obtained by averaging over the colors and polarizations of the initial state, and summing over the colors and the polarizations of the final state. This is performed using FeynCalc [36, 37]. This latter software is not really adapted to this kind of calculations, and for this reason, a long computer code has been developed to do such computations.

Ignoring quarks masses, we obtain the flowing expressions:

(a) Space-space non-commutativity:

$$\left(\frac{d\hat{\sigma}(qg \to q\gamma)}{d\hat{t}}\right)_{space-space}$$
$$= \frac{\pi q^2 \alpha_e \alpha_s(\mu_R^2)}{12288\hat{s}^3\hat{u}}(\hat{s}^2(\sin(\theta)(\sin(\varphi)(\sin(\theta)\sin(\varphi)(B_1^2\hat{s}^2\sin(\theta)\sin(\varphi)(\sin(\theta)\sin(\varphi)(B_1^2(\hat{t}^2 + \hat{t}\hat{u} + 2\hat{u}^2) + 2B_3^2\hat{u}^2) + 2B_2B_3\hat{u}(3\hat{t} + 2\hat{u})) - 2\hat{u}^2(B_1^2(4B_3^2 + 3B_{12}^2)\hat{s}^2 - 256B_{13}^2) + 256B_1^2\hat{t}^2 + 256B_1^2\hat{t}\hat{u}) + 512B_2B_3u(3\hat{t} + 2\hat{u})) + B_2^2\hat{s}^2\sin^3(\theta)\cos^4(\varphi)(B_2^2(\hat{t}^2 + \hat{t}\hat{u} + 2\hat{u}^2) + 2B_3^2\hat{u}^2) + 2B_1B_2\hat{s}^2\sin^2(\theta)\cos^3(\varphi)(2\sin(\theta)\sin(\varphi)(B_2^2(\hat{t}^2 + \hat{t}\hat{u} + 2\hat{u}^2) + B_3^2\hat{u}^2) - B_2B_3\hat{u}(3\hat{t} + 2\hat{u})) + 2\sin(\theta)\cos^2(\varphi)(\hat{s}^2\sin(\theta)\sin(\varphi)(\sin(\theta)\sin(\varphi)(B_2^2(3B_1^2(\hat{t}^2 + \hat{t}\hat{u} + 2\hat{u}^2) + B_3^2\hat{u}^2) + B_1^2B_3^2\hat{u}^2) + B_2(B_2^2 - 2B_1^2)B_3\hat{u}(3\hat{t} + 2\hat{u})) - \hat{u}^2(B_2^2(4B_3^2 + 3B_{12}^2)\hat{s}^2 - 256B_{23}^2) + 128B_2^2\hat{t}^2 + 128B_2^2\hat{t}\hat{u}) + 2B_1\cos(\varphi)(\sin(\theta)\sin(\varphi)(\hat{s}^2\sin(\theta)\sin(\varphi)(2B_2\sin(\theta)\sin(\varphi)(B_1^2(\hat{t}^2 + \hat{t}\hat{u} + 2\hat{u}^2) + B_3^2\hat{u}^2) + (2B_2^2 - B_1^2)B_3\hat{u}(3\hat{t} + 2\hat{u})) - 2B_2\hat{u}^2((4B_3^2 + 3B_{12}^2)\hat{s}^2 - 256) + 256B_2\hat{t}^2 + 256B_2\hat{t}\hat{u} - 256B_3\hat{u}(3\hat{t} + 2\hat{u}))) + 2\hat{u}\cos(\theta)(\hat{s}^2\sin^2(\theta)(B_1\sin(\varphi) + B_2\cos(\varphi))^2 + 256)(B_{12}^2(3\hat{t} + 2\hat{u}) + 2\hat{u}B_3\sin(\theta)(B_2\sin(\varphi) - B_1\cos(\varphi))) + 2B_{12}^2\hat{u}^2\cos^2(\theta)(\hat{s}^2\sin^2(\theta)(B_1\sin(\varphi) + B_2\cos(\varphi))^2 + 256)) + 512(\hat{s}^2(-((4B_3^2 + 3B_{12}^2)\hat{u}^2 + 8)) - 8\hat{u}^2))$$

(A.10)

$$\left(\frac{d\hat{\sigma}(q\bar{q} \to g\gamma)}{d\hat{t}}\right)_{space-space}$$
$$= \frac{\pi q^2 \alpha_e \alpha_s(\mu_R^2)}{4608\hat{s}^2\hat{t}\hat{u}}(2048(\hat{t}^2(\hat{u}^2B^2 + 2) + 2\hat{u}^2) + B_1^4\hat{s}^4(\hat{t}^2 + \hat{u}^2)\sin^4(\theta)\sin^4(\varphi) + B_2^4\hat{s}^4(\hat{t}^2 + \hat{u}^2)\sin^4(\theta)\cos^4(\varphi) + 4B_1B_2^3\hat{s}^4(\hat{t}^2 + \hat{u}^2)\sin^4(\theta)\sin(\varphi)\cos^3(\varphi) + 128\hat{s}^2\sin^2(\theta)\sin^2(\varphi)(B_1^2(\hat{t}^2 + \hat{u}^2) - 4B_3^2\hat{t}\hat{u}) + 2\hat{s}^2\sin^2(\theta)\cos^2(\varphi)(3B_1^2B_2^2\hat{s}^2(\hat{t}^2 + \hat{u}^2)\sin^2(\theta)\sin^2(\varphi) + 64(B_2^2(\hat{t}^2 + \hat{u}^2) - 4B_3^2\hat{t}\hat{u})) - 512\hat{t}\hat{u}B_2B_3\sin(\varphi)(\hat{s}^2\sin(2\theta) + 2(\hat{t} - \hat{u})(\hat{t} + \hat{u})\sin(\theta)) - 512B_{12}^2\hat{t}\hat{u}\hat{s}^2\cos^2(\theta) + 4B_1\cos(\varphi)(B_1^2B_2\hat{s}^4(\hat{t}^2 + \hat{u}^2)\sin^4(\theta)\sin^3(\varphi) + 64B_2\hat{s}^2(\hat{t}^2 + \hat{u}^2)\sin^2(\theta)\sin(\varphi) + 128B_3\hat{t}\hat{u}(\hat{s}^2\sin(2\theta) + 2(\hat{t} - \hat{u})(\hat{t} + \hat{u})\sin(\theta))) + 1024B_{12}^2\hat{t}\hat{u}(\hat{u}^2 - \hat{t}^2)\cos(\theta))$$

(A.11)

(b) Space-time non-commutativity:

$$\left(\frac{d\hat{\sigma}(qg \to q\gamma)}{d\hat{t}}\right)_{space-time}$$
$$= -\frac{\pi q^2 \alpha_e \alpha_s(\mu_R^2)}{12288\hat{s}^3\hat{u}}(\hat{s}^5(-E_3^4)(\hat{t} + 2\hat{u}) + 2\hat{s}^4E_3^2(E_{12}^2\hat{t}\hat{u} + 128) + \hat{s}^2(\sin(\theta)(E_1\cos(\varphi) + E_2\sin(\varphi)) + E_3\cos(\theta))((\sin(\theta)(E_1\cos(\varphi) + E_2\sin(\varphi)) + E_3\cos(\theta))(\hat{s}^2(\sin(\theta)(E_1\cos(\varphi) + E_2\sin(\varphi)) + E_3\cos(\theta))((\hat{t}^2 + \hat{t}\hat{u} + 2\hat{u}^2)(\sin(\theta)(E_1\cos(\varphi) + E_2\sin(\varphi)) + E_3\cos(\theta)) + 2E_3(2\hat{t}^2 + 5\hat{t}\hat{u} + 4\hat{u}^2)) + 2(3\hat{s}^2E_3^2(\hat{s}^2 + \hat{u}^2) + s^2E_{12}^2\hat{t}\hat{u} + 64\hat{t}^2 + 128\hat{u}^2)) + 2E_3(\hat{s}^2(E_3^2(2\hat{t}^2 + 3\hat{t}\hat{u} + 4\hat{u}^2) - 2E_{12}^2\hat{t}\hat{u} + 128) + 128\hat{u}^2)) - 128s^2(E_3^2\hat{t}^2 - 4E_{12}^2\hat{t}\hat{u} - 32) + 4096\hat{u}^2)$$

(A.12)

$$\left(\frac{d\hat{\sigma}(q\bar{q} \to g\gamma)}{d\hat{t}}\right)_{space-time}$$
$$= \frac{\pi q^2 \alpha_e \alpha_s(\mu_R^2)}{294912\hat{s}^2\hat{t}\hat{u}}(-32\hat{t}^4(3\hat{s}^2 E_3^2(2E_3^2 + E_{12}^2) + 64(4E_3^2$$
$$+ 3E_{12}^2)) + (\hat{t}^2 + \hat{u}^2)(-16\hat{s}^2(E_1 - E_2)(E_1$$
$$+ E_2)\sin^2(\theta)\cos(2\varphi)(\hat{s}^2(E_{12}^2 - 6C03^2)\cos(2\theta)$$
$$+ \hat{s}^2(-(E_{12}^2 - 6E_3^2)) + 768)$$
$$- 32\hat{s}^2 E_1 E_2 \sin^2(\theta)\sin(2\varphi)(\hat{s}^2(E_{12}^2 - 6E_3^2)\cos(2\theta)$$
$$+ \hat{s}^2(-(E_{12}^2 - 6E_3^2)) + 768) - 4\hat{s}^2\cos(2\theta)(\hat{s}^2(40E_3^4$$
$$- 24E_3^2 E_{12}^2 + 3E_{12}^{2^2}) - 1536(E_{12}^2 - 2E_3^2)) + 8\hat{s}^4(E_1^4$$
$$- 6E_1^2 E_2^2 + E_2^4)\sin^4(\theta)\cos(4\varphi) + 32\hat{s}^4 E_1 E_2(E_1$$
$$- E_2)(E_1 + E_2)\sin^4(\theta)\sin(4\varphi) + \hat{s}^4(8E_3^4 - 24E_3^2 E_{12}^2$$
$$+ 3E_{12}^{2^2})\cos(4\theta) + 512E_1 E_3(\hat{t} - \hat{u})\sin(\theta)\cos(\varphi)(\hat{s}^3 E_3^2 \quad (A.13)$$
$$+ 64\hat{s}) + 512E_2 E_3(\hat{t} - \hat{u})\sin(\theta)\sin(\varphi)(\hat{s}^3 E_3^2 + 64\hat{s})$$
$$+ 16E_3(\hat{s}^2\sin(2\theta)(E_1\cos(\varphi) + E_2\sin(\varphi))(\hat{s}^2(4E_3^2$$
$$- 3E_{12}^2)\cos(2\theta) + 4E_3^2(\hat{s}^2 - 6\hat{s}^2) + 3(E_{12}^2\hat{s}^2 - 512))$$
$$+ 4\cos(\theta)(\hat{s}^4\sin^3(\theta)(E_1(E_1^2 - 3E_2^2)\cos(3\varphi) - E_2(E_2^2$$
$$- 3E_1^2)\sin(3\varphi)) + 8E_3(\hat{t} - \hat{u})(\hat{s}^3 E_3^2 + 64\hat{s}))))$$
$$+ 64\hat{t}^3\hat{u}(\hat{s}^2 E_3^2(10E_3^2 + 21E_{12}^2) + 192(4E_3^2 + 7E_{12}^2))$$
$$+ 64\hat{t}\hat{u}^3(\hat{s}^2 E_3^2(10E_3^2 + 21E_{12}^2) + 192(4E_3^2 + 7E_{12}^2))$$
$$+ \hat{u}^2(9\hat{s}^4 E_{12}^{2^2} + 8E_3^2(3\hat{s}^2(E_{12}^2(\hat{s}^2 - 4\hat{u}^2) - 512)$$
$$- 1024\hat{u}^2) - 24E_3^4(7\hat{s}^4 + 8\hat{s}^2\hat{u}^2) - 6144E_{12}^2\hat{u}^2$$
$$+ 262144) + \hat{t}^2(E_{12}^2(9\hat{s}^4 E_{12}^2 + 53248\hat{u}^2)$$
$$+ 8E_3^2(\hat{s}^2(E_{12}^2(3\hat{s}^2 + 104\hat{u}^2) - 1536) - 2048\hat{u}^2)$$
$$- 24E_3^4(7\hat{s}^4 + 16\hat{s}^2\hat{u}^2) + 262144))$$

Where, $(\hat{s}, \hat{t}, \hat{u})$ indicates the standard parton-level Mandelstam variables. Using Eqs. (A.7), (A.8) and (A.9), it is easy to show that $\hat{u} = -\hat{s}/2(1 + \cos\theta)$ and $\hat{t} = \hat{s}/2(-1 + \cos\theta)$.

To obtain the total expression of the differential cross section in the laboratory frame, located at $\delta$ latitude (see Fig. 3), the components of $E_i$ and $B_i$ must be replaced by their expressions obtained by applying a series of rotations Eqs. (4) and (5).

$$\{\vec{E}, \vec{B}\} \to \{\vec{E'}, \vec{B'}\} = \frac{1}{\Lambda_{e,b}^2} \times$$
$$\begin{pmatrix} -\cos\gamma_{e,b}\cos\delta\sin\alpha + \sin\gamma_{e,b}(\cos(\beta_{e,b} - a)\sin\alpha\sin\delta - \cos\alpha\sin(\beta_{e,b} - a)) \\ \cos\gamma_{e,b}\sin\delta + \cos\delta\cos(\beta_{e,b} - a)\sin\gamma_{e,b} \\ \cos\alpha(\cos\gamma_{e,b}\cos\delta - \cos(\beta_{e,b} - a)\sin\gamma_{e,b}\sin\delta) - \sin\alpha\sin\gamma_{e,b}\sin(\beta_{e,b} - a) \end{pmatrix} \quad (A.14)$$

We indicate by β and γ the spherical coordinates of $\vec{E}$ and $\vec{B}$; the azimuthal angle and colatitude, respectively (see Fig. 3 (a)). The parameter "a" represents the right ascension of the laboratory site, where collisions take place.

Since the collisions occur for several months, we average over β , φ and the time t (therefore on a). We keep only the colatitudes γ of $\vec{E}$ and $\vec{B}$, which remains fixed during this period.

The inclusive cross section is obtained by an incoherent summation over all the possible sub-processes, as follow [35]:

$$\frac{d^2\sigma}{dp_T^\gamma d\eta} = 2p_T^\gamma \sum_{ab} \int \frac{dx_a}{x_a} F_{a/A}(x_a, \mu_F^2) \int \frac{dx_b}{x_b} F_{b/B}(x_b, \mu_F^2) \hat{s} \frac{d\hat{\sigma}(\hat{s}, \mu_R^2)}{d\hat{t}} \delta(\hat{s} + \hat{t} + \hat{u}) \quad (A.15)$$

"a" and "b" represent the partons which interact from hadrons "A" and "B". $F_{a/A}(x_a, \mu_F^2)$ and $F_{b/B}(x_b, \mu_F^2)$ represent the parton distribution functions of the initial partons "a" and "b" in the hadrons "A" and "B", with momentum fractions $x_a$ and $x_b$, respectively. $\hat{\sigma}$ is the partonic cross section.

In our study, the calculations are performed using the most recent PDF CT14 [46]. The factorization scale (μ$_F$) and the renormalisation scale (μ$_R$), at which the PDF and strong coupling are evaluated, have been set equal to the photon transverse energy : $\mu_F = \mu_R = P_T^\gamma$. For numerical evaluation of the inclusive cross section, we use Adaptive Monte Carlo Method. Some computational results are shown in Fig. 4.